\documentstyle[11pt,ysc,twoside,epsf]{article}
\def\edcomment#1{\iffalse\marginpar{\raggedright\sl#1\/}\else\relax\fi}
\reversemarginpar

\begin{document}
\title{Cosmological applications of the Szekeres model}

\author{Krzysztof  Bolejko}
\affil{Nicolaus Copernicus Astronomical Center, Polish Academy of Sciences,
 ul. Bartycka 18, 00-716 Warsaw, Poland}

\begin{abstract}
This paper presents the cosmological applications of the
quasispherical Szekeres model. The quasispherical Szekeres model is
an exact solution of the Einstein field equations, which represents
a time--dependent mass dipole superposed on a monopole and therefore
is suitable for modelling double structures such as  voids and
adjourning galaxy superclusters. Moreover, as the Szekeres model is
an exact solution of the Einstein equations it enables tracing light
and estimation of the impact of cosmic structures on light
propagation. This paper presents the evolution of a void and
adjourning supercluster and also reports on how the Szekeres model
might be employed either  for the estimation of mass of galaxies
clusters or for the estimation of the luminosity distance.
\end{abstract}

\section{Introduction}

When performing astronomical observations one has to keep in mind that  light
from observed objects propagates through the Universe which is a complicated
and evolving system.  Although the evolution of the Universe can be neglected on small
scales, it cannot be done so if distant objects, such as high--redshift
galaxies, quasars or very remote supernovae are observed. Additionally, on
large scales the impact of the cosmic structures on light propagation must be
considered. Since in the Newtonian mechanics matter does not affect light
propagation, the general relativity must be employed.

This paper shows that such problems of high--redshift astronomy can
be solved by employing the quasispherical Szekeres model (Szekeres
1975a). The quasispherical Szekeres model is an exact solution of
the Einstein field equations, which represents  a time--dependent
mass dipole superposed on a monopole and therefore it is suitable
for modelling double structures such as  voids and  adjourning
galaxy superclusters. Moreover, as the Szekeres model is an exact
solution of the Einstein equations it enables tracing light and
estimation of the impact of cosmic structures on light propagation.

The structure of this paper is as follows: Sec. 2 presents the
astronomical observations  of the Universe; Sec. 3 presents the
theoretical approach to these data; Sec. 4 presents the Szekeres
model; in Sec. 6 the evolution of a void and an adjourning cluster
within the quasispherical Szekeres model is studied; the algorithm
which was employed for these calculations is presented in Sec. 5;
Sec. 7 presents how  the quasispherical Szekeres model can be
adopted to analysis of the astronomical observations.

\section{Astronomical data}\label{astrodata}

At the end of 1970s when galaxy redshift surveys started to collect
the data, the large scale cosmic structures were discovered. It
turned out that the Universe is very inhomogeneous, and galaxies
form structures like voids, clusters, and filaments. The density
contrast within these structures varies from $\delta \approx -0.94$
in voids (Hoyle \& Vogeley 2004) to $\delta$ equal to several tens
in clusters (Bardelli et al. 2000 ). These structures are of
diameters varying  from several Mpc up to several tens of Mpc.
However, if averaging is considered on large scales, the density
varies from $\delta = -0.7$ to $\delta  = 3.4 \rho_b$ (Kolat, Dekel,
\& Lahav 1995;  Hudson 1993).

Fig. 1 presents the galaxies distribution in a slice 1600~km/s thick
and 32000~km/s square\footnote{This Figure was obtained by Charles
Hellaby from the galaxy data taken from the NASA/IPAC Extragalactic
Database: http://nedwww.ipac.caltech.edu.}. As one can see galaxies
are distributed inhomogeneously. The galaxies presented in this
pictures are up to redshifts $z \approx 0.1$. So when the redshift
of observed objects is larger then unity (like distant quasars, GRB
or high--redshift supernovae) we have to keep in mind that light of
these objects propagates though this cosmic web. Moreover, matter
distribution does influence light propagation. Therefore, if we want
to  reduce the observational  data properly, we have to know what is
the impact of inhomogeneous matter distribution  on  results of the
astronomical observations.

\begin{figure}
\plotone{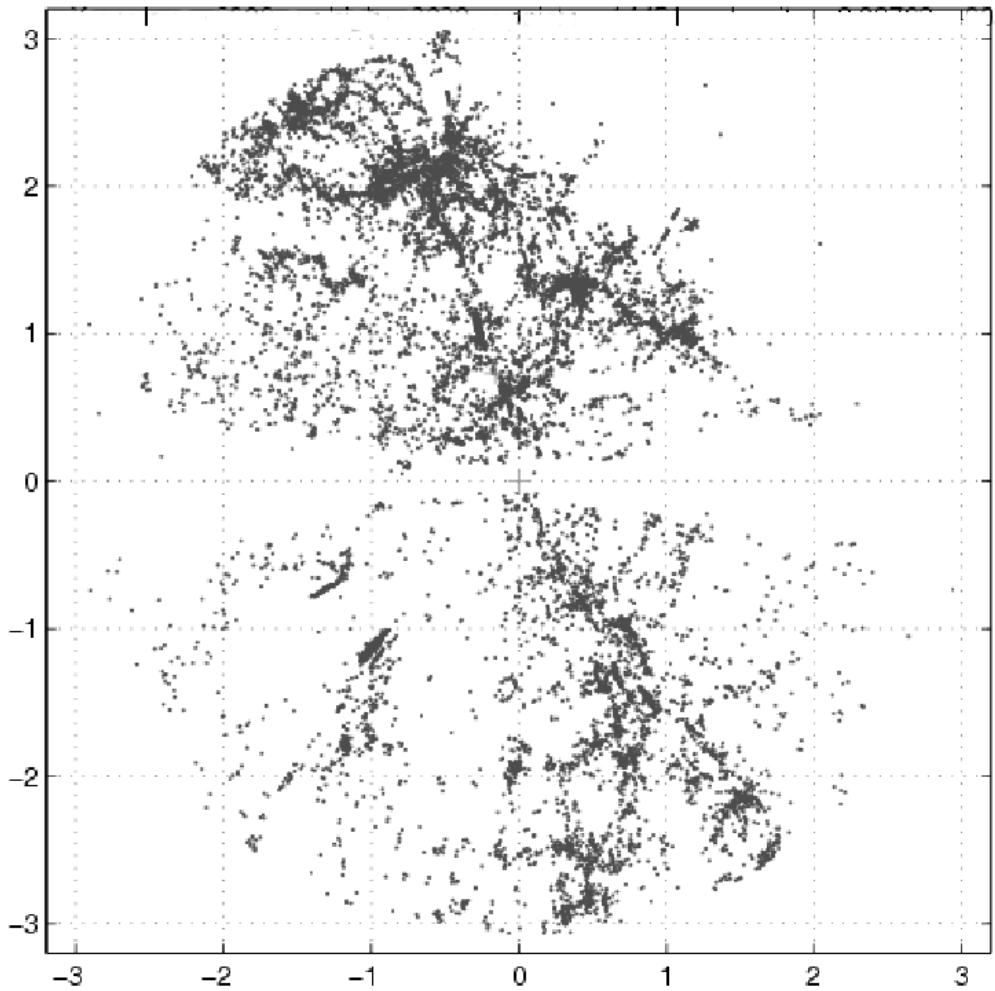} \caption{The galaxy distribution in a slice 1600~km/s thick
and 32000~km/s square. The Zone of Avoidance is clearly visible in the middle
of the picture. Since the data are from
different redshift surveys, some structures are well mapped, while other
regions are not. Still there are regions from which there is no data at all.}
\end{figure}

\section{Analysis of observations}\label{obseranaly}

As stated above, to properly analyze the observations general
relativity has to be employed. The Einstein equations of general
relativity are as follows:

\begin{equation}
G_{\alpha \beta} = \kappa T_{\alpha \beta} + \Lambda g_{\alpha \beta},
\end{equation}
where $G_{\alpha \beta}$ is the Einstein tensor which describes the
geometry of the space--time, $T_{\alpha \beta}$ is the
energy--momentum tensor, $g_{\alpha \beta}$ is the metric, and
$\Lambda$ is the cosmological constant. The Einstein equations state
that there is a correspondence between the geometry of the
space--time and matter distribution in the space--time. In general,
these equations are a set of 10 partial differential equations,
which are very hard to solve even numerically. However, if some
symmetries are assumed, these equations can be solved analytically.
The most popular assumption is that the space is homogeneous.
However, because of homogeneity one cannot describe the process of
structure formation. One of the alternatives is to employ a
perturbation theory, such as the linear approach. However, structure
formation is a very non--linear process (Bolejko 2006a; 2006b) hence
the linear approach is inadequate. Another alternative is N--body
simulation. The N--body simulation describes the evolution of a
large amount of particles which interact gravitationally. However,
the interactions between particles are described by Newtonian
mechanics, and in the Newtonian mechanics matter does not affect
light propagation. Hence within the N--body simulations it is
impossible to estimate the influence of matter distribution on light
propagation. In general relativity the situation is different, the
geometry defined by matter distribution determines  along which
paths the light will propagate. Thus to trace light propagation and
estimate the influence of cosmic structures on light propagation one
has to use an exact solution of Einstein's equations. But there are
no exact solutions which would fully describe such a complicated
system as our Universe. To describe the structure formation one has
to focus on small scales, where  one can employ models suitable for
this purpose. For evolution of single structures (such as clusters
or voids) the Lema\^itre--Tolman model (Lema\^itre 1933, Tolman
1934)  can be used, and for evolution of double structures (a void
and an adjourning cluster) the quasispherical Szekeres model can be
employed. The Szekeres model represents  a time--dependent mass
dipole superposed on a monopole. So it is suitable for modelling
double structures. The Szekeres model also makes it possible to
examine interactions between the considered structures.

\section{Szekeres model}\label{szekmdl}

For our purpose it is convenient to use a coordinate system different from
that in which Szekeres (1975a) originally found his solution. The metric is of
the following form (Hellaby \& Krasi\'nski 2002):

\begin{equation}
ds^2 =  c^2 dt^2 - \frac{(\Phi' - \Phi \frac{\textstyle E'}{\textstyle E})^2}
{(\varepsilon - k)} dr^2 - \Phi^2 \frac{(dp^2 + dq^2)}{E^2}, \label{dshk}
 \end{equation}
where ${}' \equiv \partial/\partial r$, $\varepsilon = \pm1,0$ and $k = k(r)
\leq \varepsilon$ is an arbitrary function of $r$.

The function $E$ is given by:
 \begin{equation}
E(r,p,q) = \frac{1}{2S}(p^2 + q^2) - \frac{P}{S} p - \frac{Q}{S} q + C ,
 \end{equation}
where the functions $S = S(r)$, $P = P(r)$, $Q = Q(r)$, and $C = C(r)$ satisfy
the relation:
 \begin{equation}
C = \frac{P^2}{2S} + \frac{Q^2}{2S} + \frac{S}{2} \varepsilon,~~~~~~~~~
\varepsilon = 0, \pm 1,
 \end{equation}
but are otherwise arbitrary.

As can be seen from (2), only $\varepsilon = +1$ allows the model to have all
three FLRW limits (hyperbolic, flat, and  spherical). This follows from the
requirement of the Lorentzian signature of the metric (2). As we are interested
in the Friedmann limit of our model, i.e. we expect it becomes a homogeneous
Friedmann model at a large distance from the origin, we will focus only on the
$\varepsilon = 1$ case. The $\varepsilon =1$ case is called the
quasispherical Szekeres model.

The Einstein equations reduce to the following two:

\begin{equation}
\frac{1}{c^2}\dot{\Phi}^2 = \frac{2M}{\Phi} - k + \frac{1}{3} \Lambda
\Phi^2, \label{vel}
\end{equation}

\begin{equation}
\kappa \rho c^2 = \frac{ 2 M' - 6 M E'/E}{\Phi^2 ( \Phi' - \Phi E'/E)}.
\label{rho}
\end{equation}
where $\dot{} \equiv \partial/\partial t$,  $\rho$ is matter density, and
$\kappa = 8 \pi G / c^4$.

In the Newtonian limit $c^2 M(r)/G$ is equal to the mass inside the shell of
radial coordinate $r$.  However, it is not an integrated rest mass but the
active gravitational mass that generates the gravitational field. By analogy
with the Newtonian energy conservation equation, eq. (5) shows that the
function $(- k/2)$ represents the energy per unit mass of the particles in the
shells of matter at constant $r$. On the other hand, by analogy with the
Friedman equation, and from the metric (2), the function $k$ determines the
geometry of the spatial sections $t = $const. However, since $k$ is a function
of the radial coordinate, the geometry of the space is now  position dependent.

Eq. (5) can be integrated:

\begin{equation}
\int\limits_0^{\Phi}\frac{{\rm d}
\tilde{\Phi}}{\sqrt{\frac{2M}{\tilde{\Phi}} - k + \frac{1}{3} \Lambda
\Phi^2}} = ct- ct_B, \label{cal}
\end{equation}
where $t_B = t_B(r)$ appears as an integration constant, and is an arbitrary
function of $r$. This means that the Big Bang is not a single event as in the
Friedmann models, but occurs at different times for different distances from
the origin.

As can be seen, the Szekeres model is specified by 6 functions. However, by a
choice of the coordinates, the number of independent functions can be reduced
to 5.

The Szekeres model is known to have no symmetry (Bonnor, Sulaiman, \& Tomimura
1977). It is of great flexibility and wide application in cosmology (Bonnor \&
Tomimura 1976), and in astrophysics (Szekeres 1975b; Hellaby \& Krasi\'nski
2002), and still it can be used as a model of many astronomical phenomena. This
paper aims to present the application of the quasispherical Szekeres model to the process of
structure formation.

\section{The Algorithm}\label{szapp}

Since the quasispherical Szekeres model represents a time--dependent mass
dipole superposed on a monopole, we will employ this model to describe the
evolution of a void and an adjourning galaxy supercluster. Below the algorithm
of the model's specification and its evolution is presented.

\subsection{The model setup}

To specify the model we need to know 5 functions. Let three out of these five
unknown functions be $P(r), Q(r), S(r)$. These function are assumed to be of
the following form:

\begin{eqnarray}
S &=& 140, \nonumber \\
P &=& 0, \nonumber \\
Q &=& -113 \ln (1+r).
\end{eqnarray}

The next two functions can be any two of the set $\{t_B(r), M(r), k(r)\}$, or
of any other combination of functions, from which these can be calculated. The
function $M(r)$ describes the active gravitational mass inside the $t = $
const, $r =$ const sphere. Let us take the mass distribution as presented in
Fig. 2. Fig. 2 presents not the $M(r)$ function itself, but  $M c^2/G$ (see a
comment after eq. (6)). Since the void is placed at the origin, the mass of the
model in Fig. 2 is below the background mass, however, it is compensated by
more dense regions, and at the distance of about $30$ Mpc the mass
distribution becomes as in the homogeneous background.

To define the model we need one more function. Let us assume that the bang
time function, $t_B(r)$, is constant and equal to zero. Then from eq. (7) the
function $k(z)$ can be calculated. The LHS of eq. (7) was calculated using the
64--points Gauss--Legendre quadrature. The value of $t$ on the RHS of eq. (7)
was assumed to be the present instant. Then the function $k$ was calculated by
the bisection method.

\begin{figure}
        \plotone{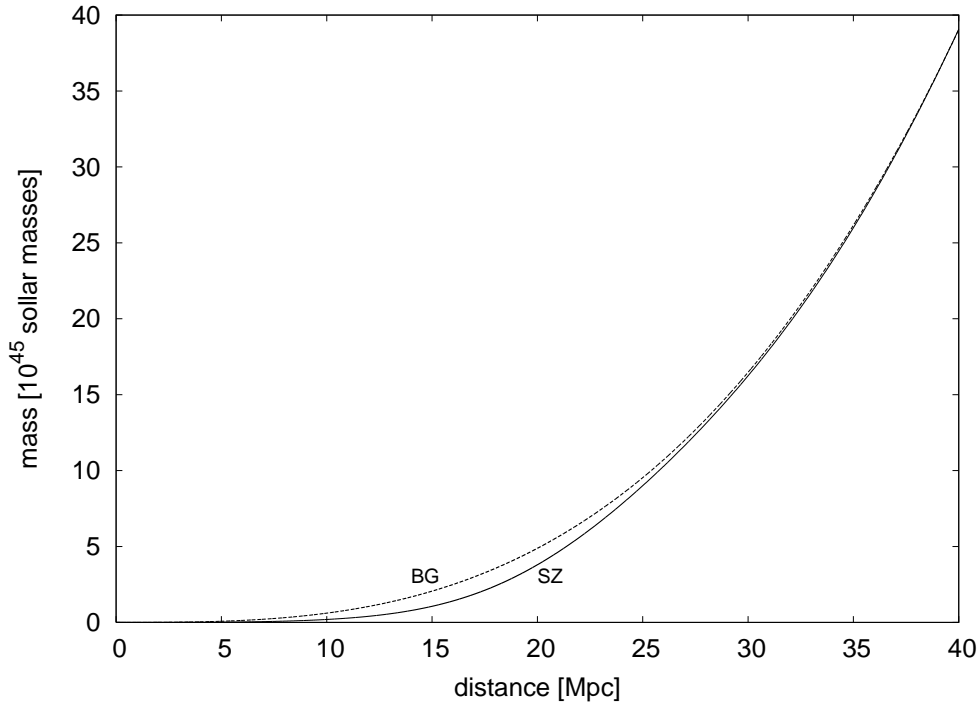}
\caption{The mass distribution within the homogeneous background (BG) and in
the Szekeres model (SZ).}
    \end{figure}

\subsection{Evolutionary code}

As can be seen  from eqs. (6), (3), and (4), to calculate the
density distribution for any instant $t_i$, one needs to know 5
functions: $M(r), S(r), Q(r), P(r)$, and $\Phi(t_i,r)$. Using these
functions the density distribution of the present day structures can
be calculated (see Sec. 6). Then, the evolution of the system can be
traced back in time. The density distribution depends on time only
via the function $\Phi(t,r)$ and its derivative. The value of
$\Phi(t,r)$ for any instant can be calculated by solving the
differential equation (5). In most cases, as in this paper, this
equation can be solved only numerically. To solve this equation one
needs to know the initial conditions: $\Phi(t_0, r)$, the functions
$M(r)$,  $k(r)$, and the value of $\Lambda$. This equation was
solved numerically using the fourth--order Runge--Kutt method.
Knowing the value of $\Phi(t,r)$ for any instant the density
distribution can be calculated as described above.

\section{Model of cosmic structures}

This section presets the evolution  of the present--day void and the
adjourning galaxy supercluster in the expanding Universe. Far away from the
origin, density and velocity distributions tend to the values that they would
have in a homogeneous Friedmann model. Fig. 3 presents the present--day
density distribution of the considered structure. On the left side of Fig. 3
the schematic cross--sections are presented, on the right side  density
distributions on these cross--sections are presented. As can be seen, at large
distance from the origin the density distribution is homogeneous and as the
distance gets smaller the structure begins to appear. At the equator the
structure is most visible. The density distribution on this equatorial
cross--section (Fig. 3, at the bottom) is also presented at the top of Fig. 4.
Fig. 4 depicts the colour coded (left side) density distributions. On the
right side of Fig. 4 the density profiles are plotted. As can be seen the
density distributions within these structures are consistent with the
observational data presented in Sec. 2.

\begin{figure}
        \plottwo{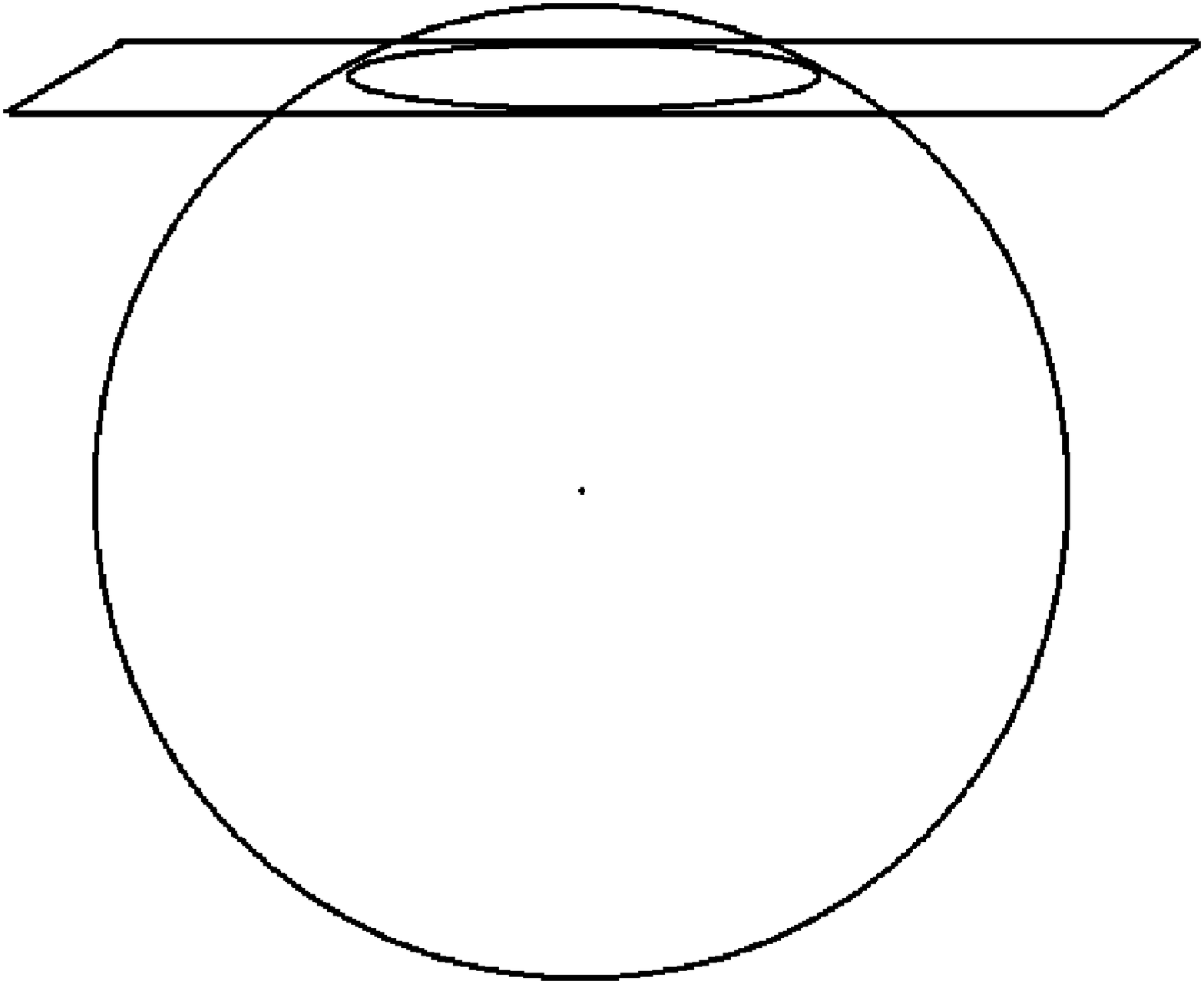}{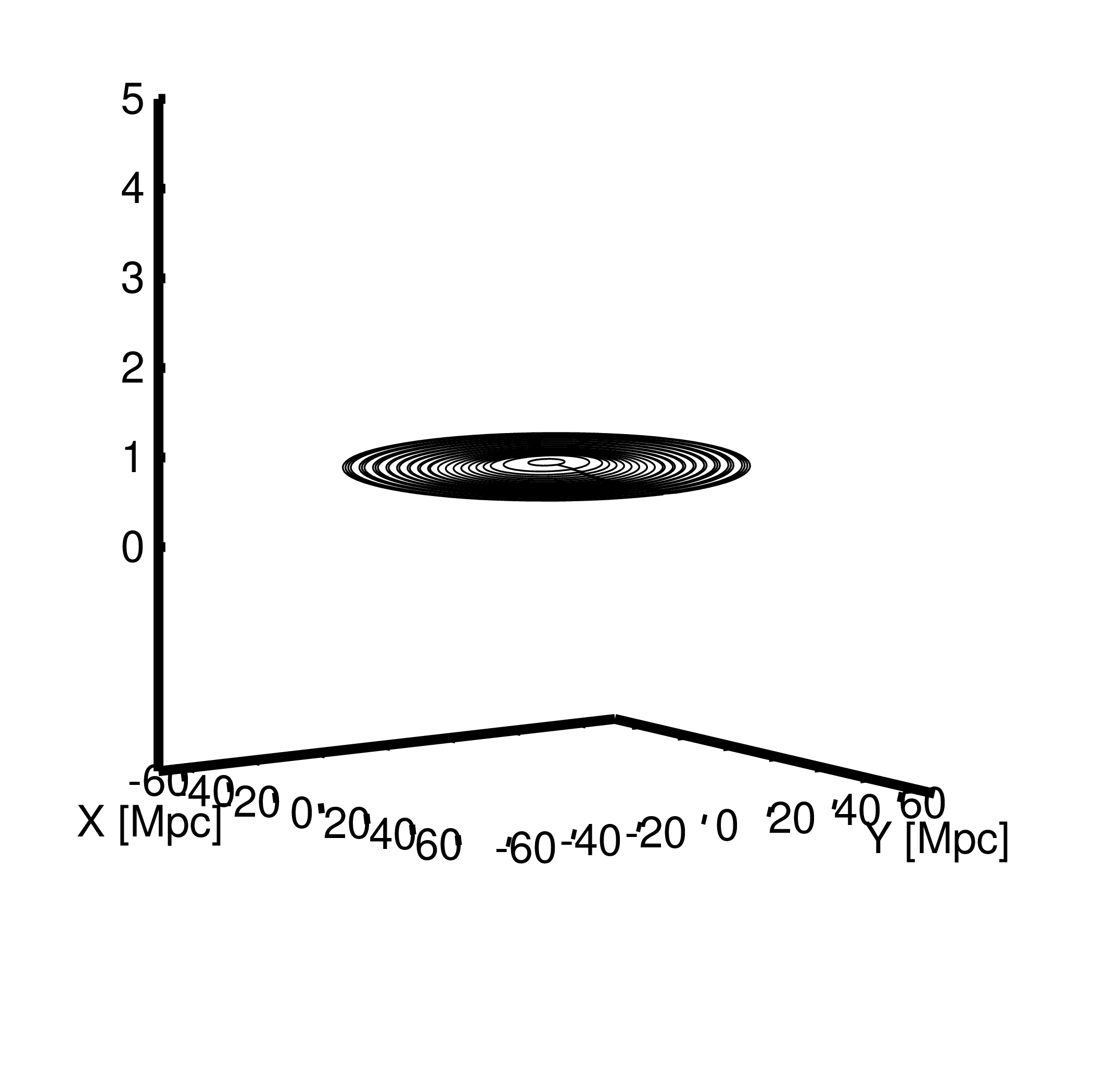}
    \plottwo{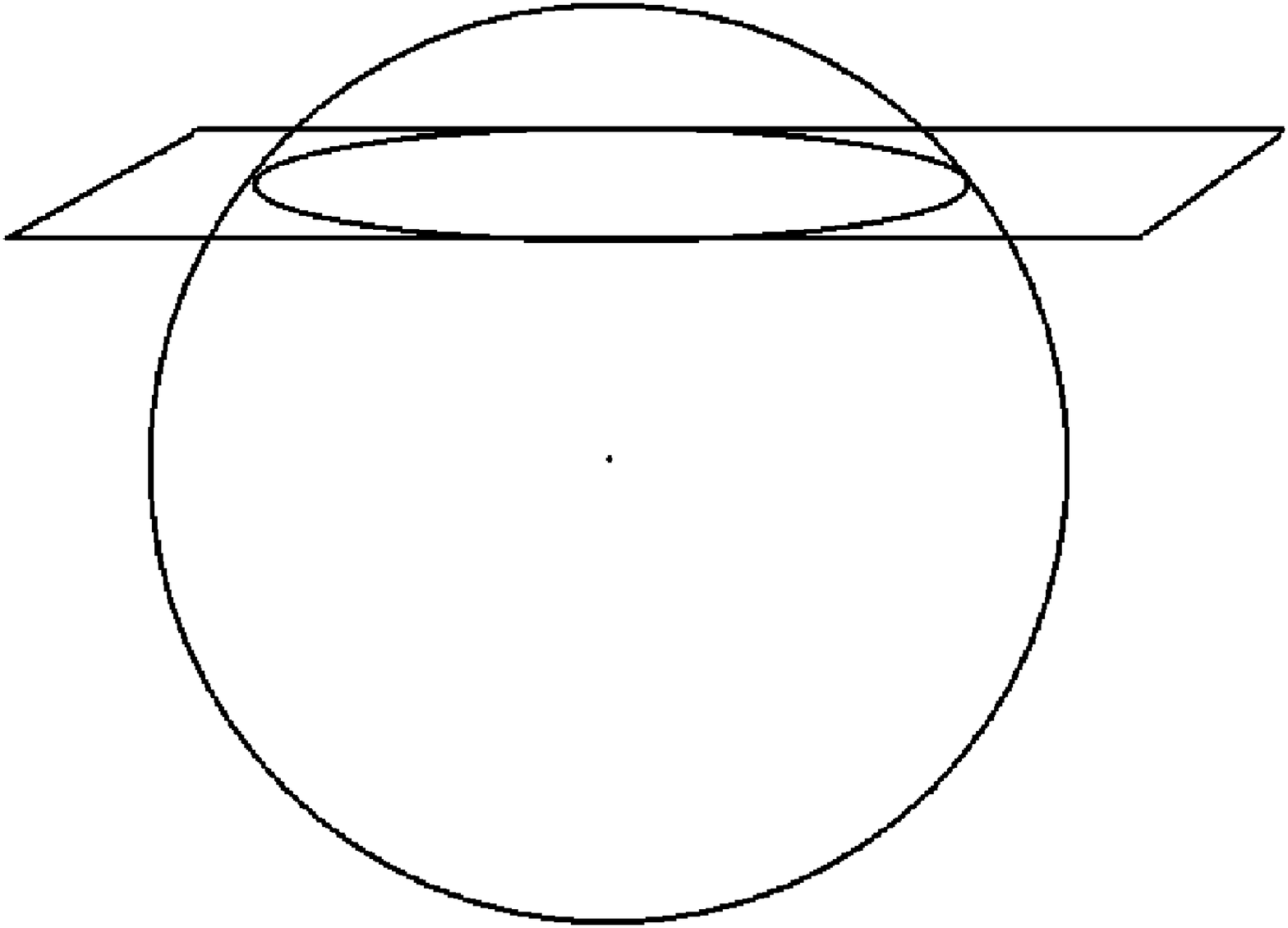}{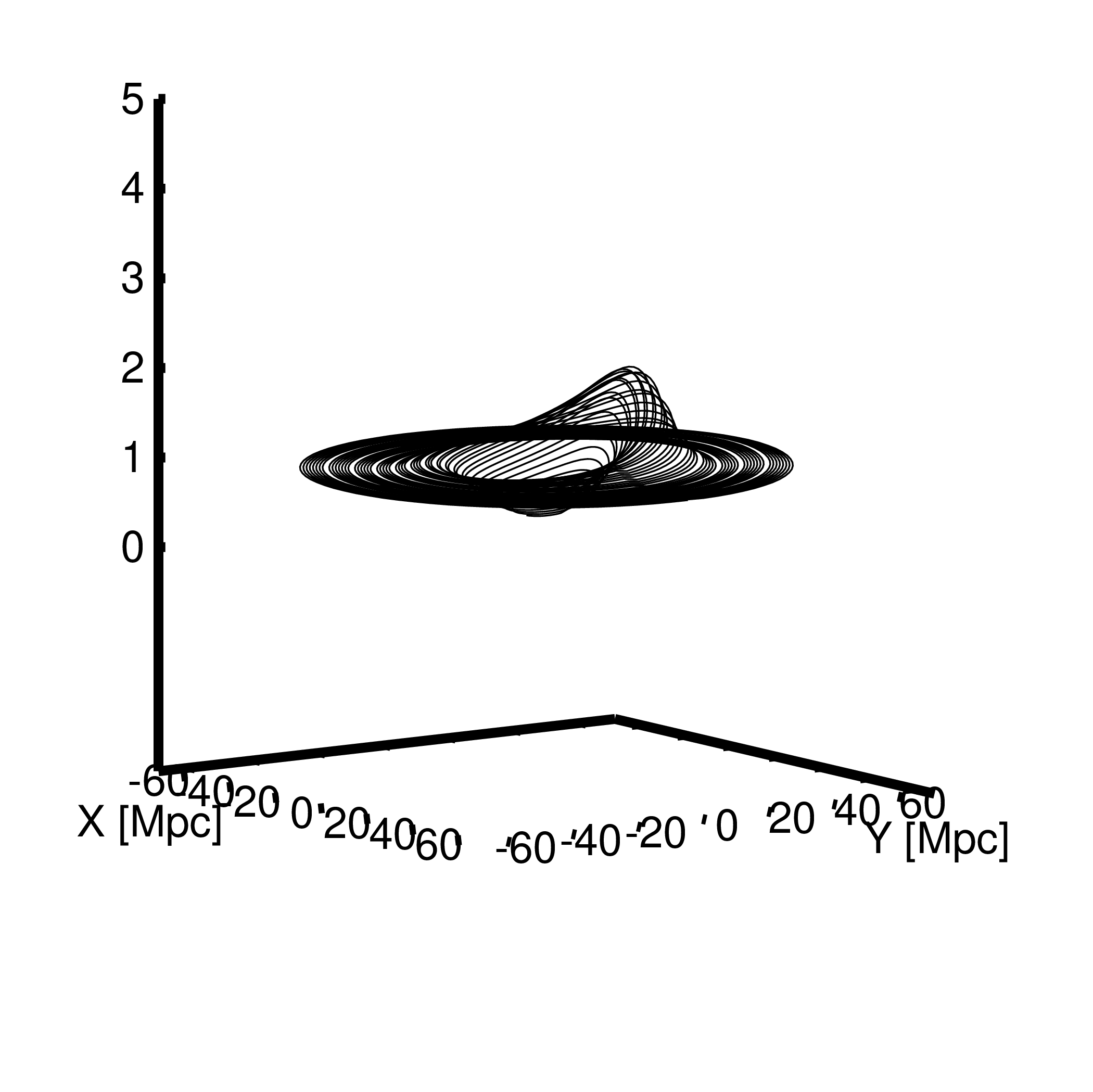}
    \plottwo{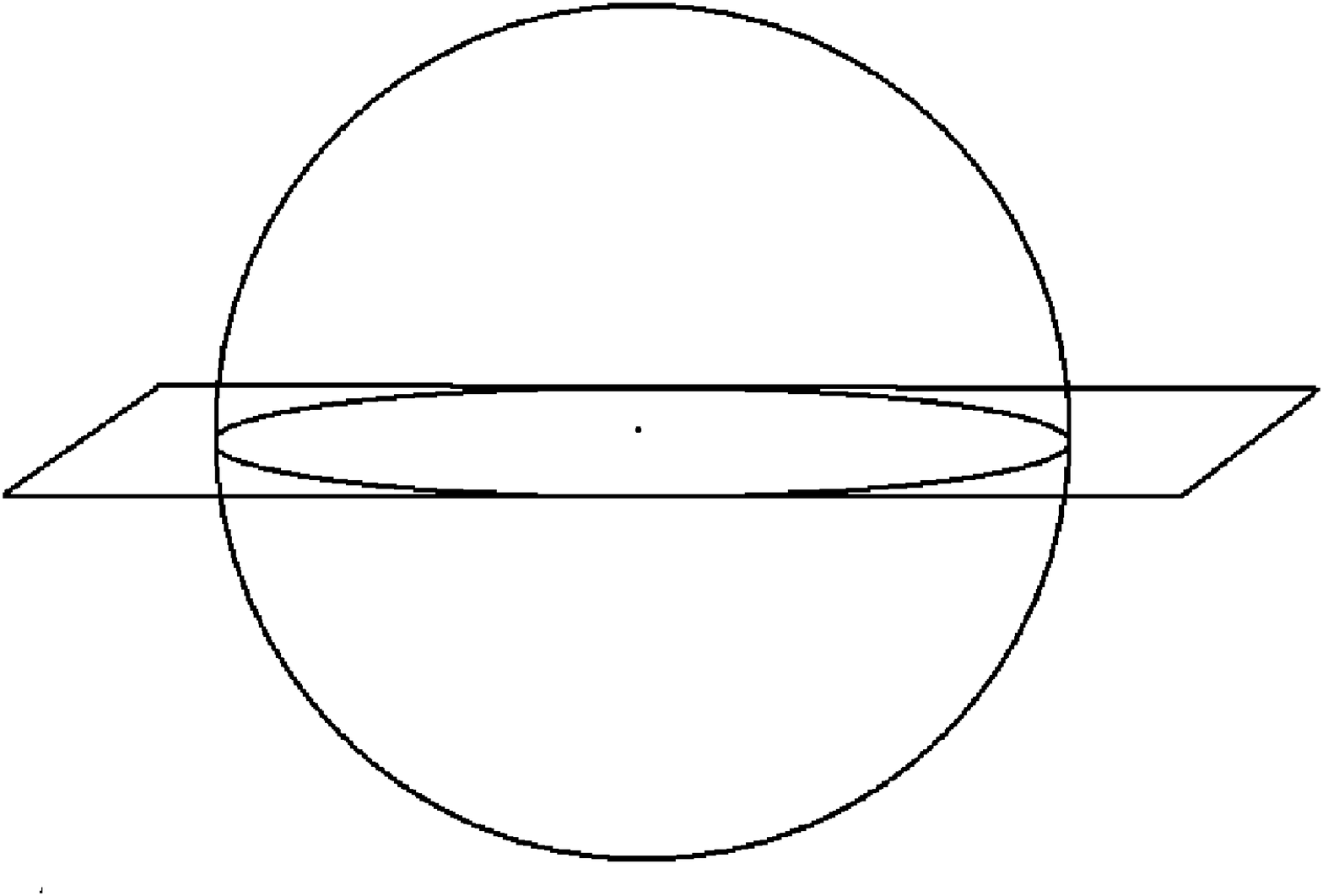}{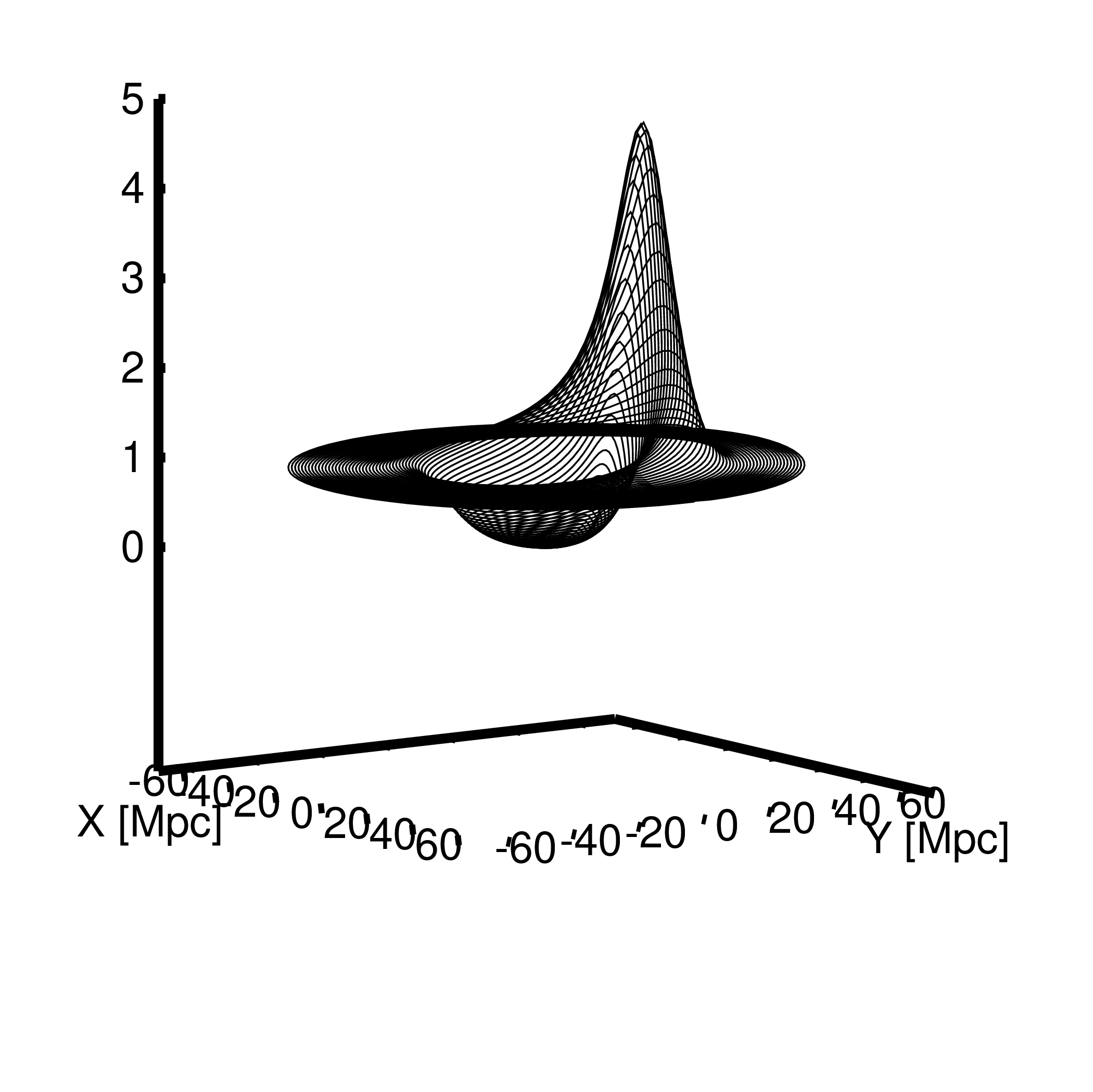}
\caption{Density distribution of the considered structure. Left side figures
present the schematic cross--sections through the modeled region. Pictures on
the right side present the density distributions which correspond to these
cross--sections. The density is presented in the background density units,
i.e. $\rho(r,t)/\rho_b (t)$, where $\rho_b$ is density of the background
model. For detailed description see Sec. 6.}
    \end{figure}

       \begin{figure}
\plottwo{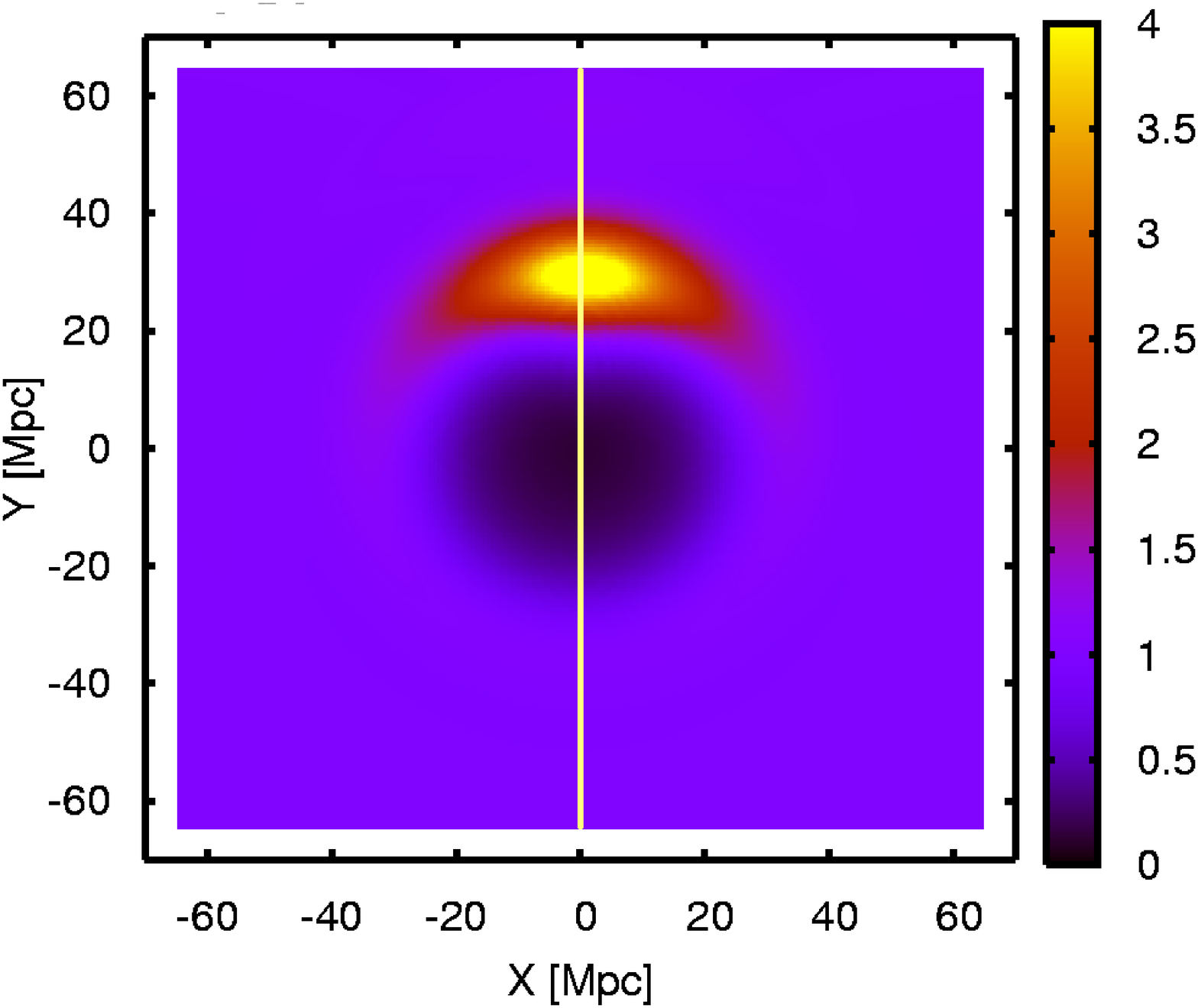}{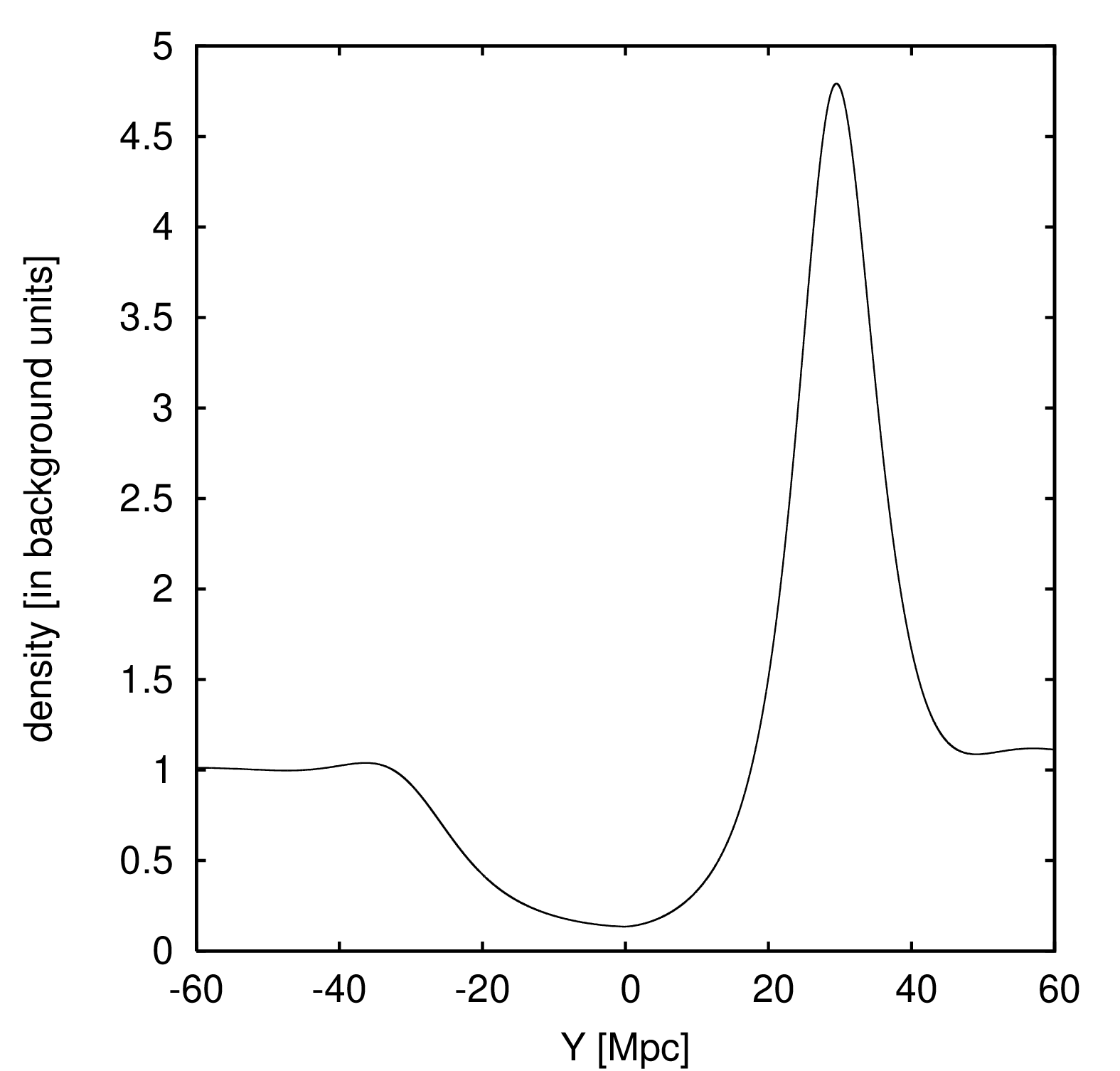} \plottwo{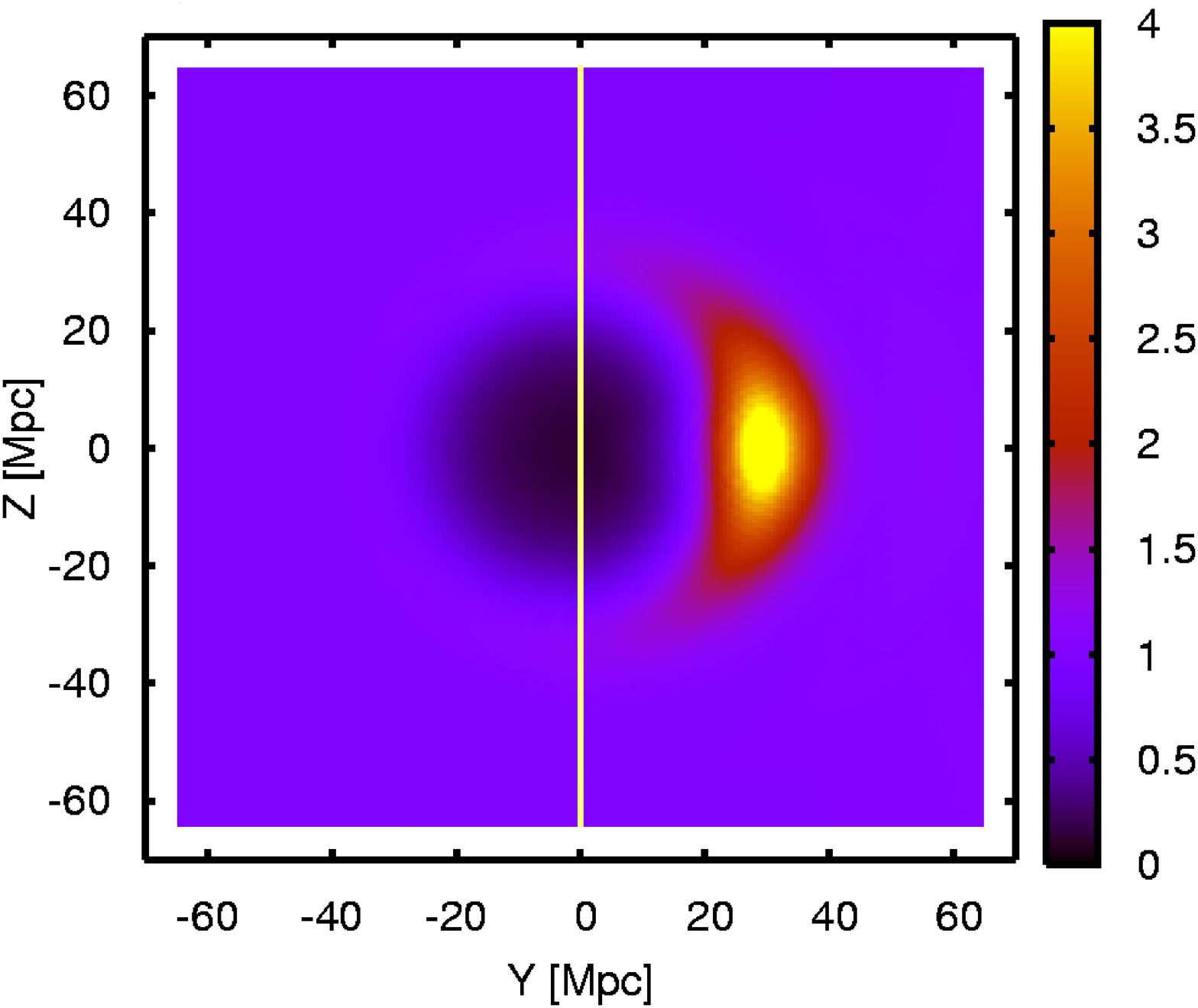}{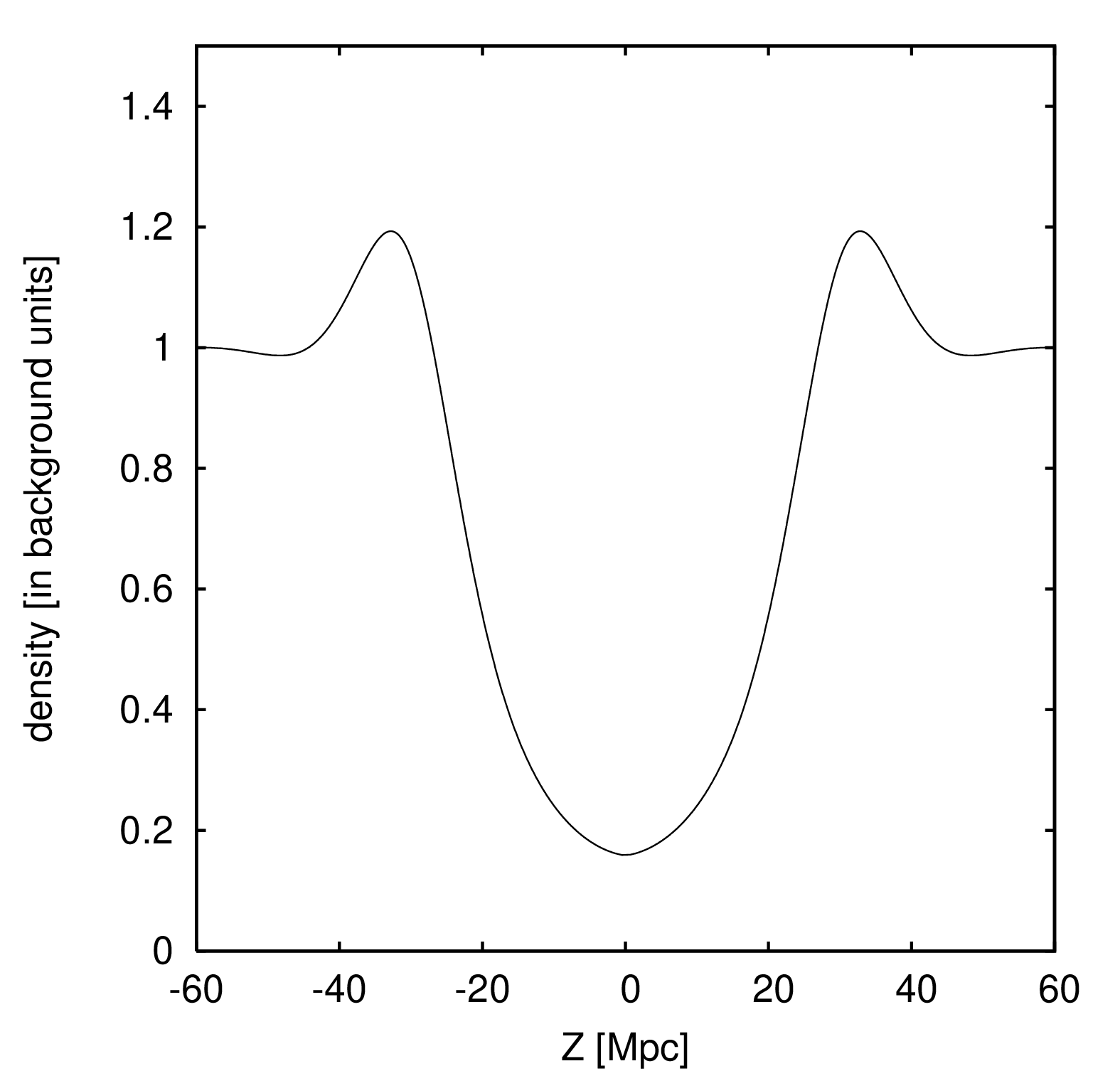}
 \caption{Density
distribution in the considered structure. Upper left panel presents colour
coded density distribution of the equatorial cross section (see Fig. 3, bottom
panels). Lower left panel presents the vertical cross--section of $X = 0 $,
through the considered model. The yellow lines correspond to the density
profiles, which are presented on the right side. For detailed description see
Sec. 6.}
    \end{figure}

\subsection{Evolution of cosmic structures}

This subsection discusses the evolution of the density profile
presented in the upper right panel of Fig. 4. In Fig. 5 density
profiles are plotted for different time instants. As can be seen in
Fig. 5 the structure formation is a non--linear process. In the
linear approach, during the evolution the shape of the initial
fluctuation does not change. Only the amplitude changes. Moreover,
in the linear approach the evolution of the density contrast does
not depend on its sign. Here, as can be seen in Fig. 5., when the
age of the Universe was 1 Gyr, the absolute value of the density
contrast inside the void was larger than inside the cluster. When
the age of the Universe was 5.5 Gyr, the absolute values of these
density contrast were comparable. Since that instant the amplitude
of the density contrast inside the cluster started to grow much
faster than the absolute value of the amplitude inside the void. For
detailed comparisons between the evolution in the linear approach
and in the quasispherical Szekeres model see Bolejko (2006b).

      \begin{figure}
\plotone{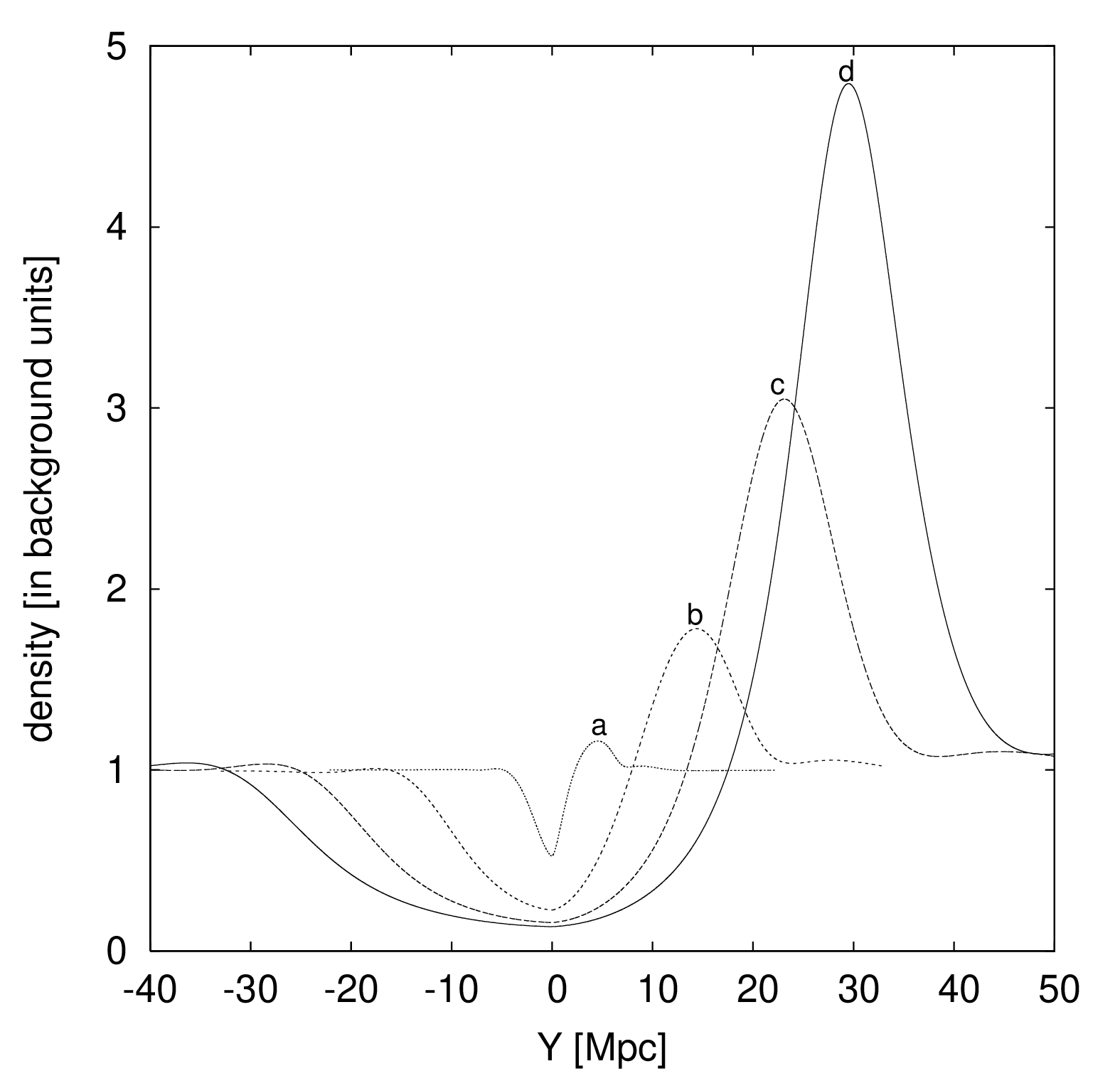}
\caption{The density profile for diffrent time instants: a --- 1 Gy after
the Big Bang, b --- 5.5 Gy, c --- 10 Gy, d --- present instant.}
    \end{figure}

\section{Futher applications of the Szekeres model}\label{fthapp}

Light propagation within the Szekeres model can be investigated by
calculating  null geodesics. Systematic studies of this problem will provide
insight in the following issues:

\begin{enumerate}
\item
The impact of matter inhomogeneities on the luminosity distance. \\
The studies of this issue within the Lema\^itre--Tolman model (Bolejko 2005)
proved that realistic matter fluctuations can change  the observed luminosity even by
$\Delta m \approx 0.15$ mag. However, since the Lema\^itre--Tolman model
assumes spherical symmetry, similar analysis in non--symmetrical models should
be repeated.
\item
Estimation of the age of the Universe. \\
The estimated age of the Universe up to the last scattering moment within the
Szekeres model and within the FLRW model is similar. This is due to homogeneous
matter distribution. However, after the last scattering when matter
inhomogeneities started to grow the results obtained within these two models
might differ. For example the time that the light will need to propagate from
regions of redshift $z \approx 1100$ is different in a homogeneous than  in an
inhomogeneous model. The difference is due to the interaction between the
cosmic structures and the photons propagating through them. This difference
can be estimated by employing the quasispherical Szekeres model.
\item
The mass of galaxy clusters. \\
The mass of a galaxy cluster can be estimated on the basis of dynamics of the
observed galaxies. The observables of this method are: the angular distance
from the center of the cluster, and the redshift. As mentioned above the
distance to the point of redshift $z$ is different in various cosmological
models. Also the angular distance from the center depends on the model. In most
cases it is assumed that an angular distance is  as in the Euclidean space ---
proportional to the physical distance. In general relativity, the path of
light of the observed galaxy can be bent by the gravitational field of the
cluster. The quasispherical Szekeres model can be used to estimate the impact of the
inhomogeneous and non--symmetrical mass distribution on the calculated mass of
observed cluster.
\end{enumerate}

\section{Conclusion}

This paper presents the cosmological  application of the
quasispherical Szekeres model. This model is  an exact solution of
the Einstein field equations, which represents a time--dependent
mass dipole superposed on a monopole. Therefore, the Szekeres model
is suitable for modelling double structures such as  voids and
adjourning galaxy superclusters. The models based on the Szekeres
solution have also one more advantage --- they can be employed in
solving problems of light propagation, which is impossible within
the N--body simulations. The Szekeres model has a great, and so far
unused, potential for applications in cosmology. It is  not only
suitable for studying the interactions between cosmic structures,
but can also  be used for estimation of the impact of matter
inhomogeneities on light propagation. The Szekeres model is suitable
for the  investigation of following issues:

--- the mass estimation based of the dynamics of galaxies,

--- the luminosity of distant objects, such as high--redshift supernovae,

--- the age of the Universe.

\section{Acknowledgments}

I would like to thank Andrzej Krasi\'nski and Charles Hellaby for
their valuable comments and discussions concerning the Szekeres
model.

\end{document}